\documentclass[amsmath,twocolumn,prx,aps]{revtex4-1} %nofootinbib
\usepackage{amsthm,amsfonts,graphicx,verbatim, color}
\usepackage{bm}% bold math
\usepackage[utf8]{inputenc}
\let\oldmarginpar\marginpar
\renewcommand\marginpar[1]{\-\oldmarginpar[\raggedleft\footnotesize #1]%
{\raggedright\footnotesize #1}}

\newcommand{\be}{\begin{equation}}
\newcommand{\ee}{\end{equation}}
\newcommand{\bea}{\begin{eqnarray}}
\newcommand{\eea}{\end{eqnarray}}

\renewcommand{\epsilon}{\varepsilon}
\renewcommand{\vec}[1]{{\bf #1}}

\newcommand{\hil}{\mathcal{D}}

\def\beq{\begin{equation}}
\def\eeq{\end{equation}}
\def\bea{\begin{eqnarray}}
\def\eea{\end{eqnarray}}
\RequirePackage[normalem]{ulem} %DIF PREAMBLE
\RequirePackage{color}\definecolor{RED}{rgb}{1,0,0}\definecolor{BLUE}{rgb}{0,0,1} %DIF PREAMBLE
\begin{document}

\title{Emergent local integrals of motion without a complete set of localized eigenstates}

\author{Scott D. Geraedts$^1$, R.N. Bhatt$^1$ and Rahul Nandkishore$^{2,3}$}
\affiliation{$^1$Department of Electrical Engineering, Princeton University, Princeton NJ 08544, USA}
\affiliation{$^2$Department of Physics and Center for Theory of Quantum Matter, University of Colorado, Boulder, Colorado 80309, USA}
\begin{abstract}
Systems where all energy eigenstates are localized are known to display an emergent local integrability, in the sense that one can construct an extensive number of operators that commute with the Hamiltonian and are localized in real space. Here we show that emergent local integrability does not require a complete set of localized eigenstates. Given a set of localized eigenstates comprising a nonzero fraction $(1-f)$ of the full many body spectrum, one can construct an extensive number of integrals of motion which are local in the sense that they have {\it nonzero weight} in a compact region of real space, in the thermodynamic limit. However, these modified integrals of motion have a `global dressing' whose weight vanishes as $\sim f$ as $f \rightarrow 0$. In this sense, the existence of a {\it non-zero fraction} of localized eigenstates is sufficient for emergent local integrability.  We discuss the implications of our findings for systems where the spectrum contains delocalized states, for systems with projected Hilbert spaces, and for the robustness of quantum integrability. 
\end{abstract}
\maketitle

\section{Introduction}

The dynamical behavior of quantum many body systems is a great open frontier for theoretical physics \cite{ReichlLin, Reichl, DynamicsReview}, which is becoming of greater importance with advances in our ability to synthesize, control and measure quantum many body dynamics in the laboratory \cite{QS1, QS2, QS3, QS4, QS5}. {\it Integrable} systems \cite{KAM1, KAM2, KAM3, KAM4, KAM5}, for which the dynamics can be calculated exactly, represent a powerful tool. However, it remains unclear how robust integrability in quantum systems is to arbitrary weak perturbations - despite decades of effort, there is no quantum analog of the Kolmogorov-Arnold-Moser (KAM) theorem of classical mechanics \cite{Kolmogorov, Arnold, Moser, Tabor}. Very recently a new paradigm has emerged in our understanding of quantum dynamics with the advent of {\it many body localization} (MBL) \cite{Anderson1958, Fleishman, AGKL, BAA, Gornyi, OganesyanHuse, Pal, Prosen, Imbrie, Nandkishore-2015}.  

MBL is a non-equilibrium phenomenon wherein quantum many body systems with strong randomness can display ergodicity breaking, failing to reach thermal equilibrium even at infinite times. Many body localized systems are a cornucopia of exotic physics, exhibiting quantum orders that cannot exist in thermal equilibrium\cite{LPQO, Pekkeretal, VoskAltman, Bauer, Bahri, ChandranSPT}, and supporting an unusual phenomenology including a strictly vanishing DC conductivity even non-zero energy densities \cite{BAA, Gornyi}, a non-local response to local perturbations \cite{nonlocal},  unusual scaling of response functions \cite{response}, and a rich structure to the pattern of quantum entanglement \cite{bardarson, geraedts}. Intriguingly, MBL also provides glimmers of hope for a quantum KAM theorem, through the notion of {\it emergent integrability}  \cite{lbits, Serbynlbits, Ros, Imbrie}.

 Emergent integrability holds that in systems where {\it all} the many-body eigenstates are localized (sometimes referred to as full-MBL or FMBL), one can construct an extensive number of operators (variously known as localized bits, lbits, local integrals of motion or LIOMs) which commute with the Hamiltonian and which are localized in real space, in the sense that the {\it weight} of the operator falls off exponentially with distance $R$ from a given point in real space. Perturbing the Hamiltonian changes the detailed structure of these LIOMs, but not the {\it existence} of an extensive number of LIOMs, such that the `emergent integrability' of the system is robust against arbitrary local perturbations. This concept of emergent integrability has rapidly become central to our understanding of MBL. However, the notion of emergent integrability is currently restricted to the FMBL regime, a restriction that must be relaxed if the notion is to impact our broader understanding of quantum dynamics.  

In this work we extend the notion of emergent integrability beyond the FMBL regime. The central problem we discuss is whether a notion of emergent integrability can be defined given a set of localized eigenstates that does {\it not} form a complete basis for the Hilbert space (i.e. a proper subset of the full set of energy eigenstates). We find that if the fraction of the Hilbert space not spanned by the set of localized states vanishes in the thermodynamic limit (i.e. the number of `missing' states grows more slowly than $d^N$, where $N$ is the number of degrees of freedom and $d$ is the dimension of the local Hilbert space), then an extensive number of LIOMs can be defined.  When the fraction of Hilbert space $f$ not spanned by the set of localized states is non-zero in the thermodynamic limit, then one can construct an extensive number of quasi local integrals of motion, which commute with the Hamiltonian and are localized in the sense that they have {\it nonzero weight} in a compact region of real space, as we take the thermodynamic limit on the size of the system. However, these modified lbits have a `global dressing' whose total weight vanishes as $\sim f$ as $f \rightarrow 0$. 

There is an obvious analogy here to Fermi liquid theory and the non-zero quasiparticle residue, in that the existence of a {\it non-zero fraction} of localized eigenstates is sufficient for emergent integrability, but with quasi-local integrals of motion that have non-zero `residue' (i.e. weight in a compact region of space). We further note that while localization of eigenstates and localization of dynamics are in principle independent concepts \cite{NGH, JNB, lstarbits}, the existence of quasi-LIOMs is sufficient for localized dynamics, since at least some memory of the initial condition will be preserved in local observables for all times, which we view as the `fundamental' definition of a localized phase \cite{Nandkishore-2015}. 
 
While the problem tackled by our work is self contained, our results also have implications for a number of other interesting problems. One example is systems where the many body spectrum contains regions of localized and delocalized states, separated by many body mobility edges. Many body mobility edges are almost universally found in perturbative and numerical studies at intermediate disorder strength. It is not clear how the picture of emergent integrability, so central to our understanding of FMBL, applies to systems with mobility edges. Our method of removing (delocalized) states from the Hilbert space is a way to attack the problem of mobility edges from a different angle. %Since we are not using any information about the projected out states, our results would be unchanged even if the states were extended, and suggests that such systems have an emergent integrability, with at least quasi local integrals of motion. 
Other applications of our work include systems where we are interested only in states in a certain energy window (e.g. a low energy subspace), and gauge theories, where gauge constraints project out certain states.

Finally, our work calls into question the strength of recent no go theorems \cite{pottervishwanath, pottervasseur} arguing for the incompatibility of MBL with certain types of ordering, or with non-Abelian symmetries. These works rely on establishing the incompatibility of strictly local integrals of motion with e.g. non-Abelian symmetries, and do not address the possibility of weaker forms of emergent integrability, such as those discussed in this work. It would be interesting to revisit these no-go theorems in light of the present work, to ask whether non-Abelian symmetries or FMBL-incompatible orders could be compatible with weaker forms of MBL with quasi local integrals of motion. 

The remainder of this paper is organized as follows. 
In Sec~\ref{sec::model} we describe the model we study and how we project into an incomplete set of eigenstates. We provide some justification as to why such a system may be described with local integrals of motion. 
In Sec.\ref{sec::Chandran} we present numerical results using a method adapted from Ref.\cite{Chandran} to construct extensively many local integrals of motion from a localized proper subset of the energy spectrum.
We discuss the implications of our work further in Sec.\ref{discussion}. 
Appendix\ref{middle} contains additional data showing that our results do not depend on which incomplete set of eigenstates is chosen, while in Appendix~\ref{bubbles} we address the recent claims of Ref.~\onlinecite{bubbles}, which call into question the existence of mobility edges in the thermodynamic limit, notwithstanding the numerous perturbative and numerical investigations that appear to see this phenomenon. Additional numerical data is presented in Appendix C. 

\section{Model and Motivation}
\label{sec::model}

The model we investigate numerically is the spin half Heisenberg spin chain with random uniaxial field
\begin{equation}
H = \sum_{i=1}^N \vec{S}_i \cdot \vec{S}_{i+1} + h_i S_i^z \label{heisenberg}
\end{equation}
where $h_i$ is taken from a bounded uniform distribution $[-h, h]$. We employ periodic boundary conditions. This model is known to have a localization transition at a critical $h$ which depends on energy density. We note that finding local integrals of motion is more numerically expensive than calculating eigenstates, so we are limited to systems with a maximum of $N=14$ spins. For this system size, all eigenstates are localized for $h_c \ge 3.5$ \cite{Pal}. 
We will be projecting out a fraction $f$ of the Hilbert space which contains an incomplete set of eigenstates.
This projection is accomplished with the operator $P\equiv V V^\dagger$, where $V$ a matrix with $(1-f) \hil$ columns, $\hil=2^N$ being the dimension of the Hilbert space.

The system with states projected out is governed by an effective Hamiltonian that is non-local, but (as we will show) retains an emergent local integrability.  In the main text we project out states at one end of the spectrum (the `top states'). Qualitatively similar results, but with more severe finite size effects, are obtained when the states projected out are in the middle of the many body spectrum (see Appendix \ref{middle}). 

\begin{figure}
\includegraphics[width=\linewidth]{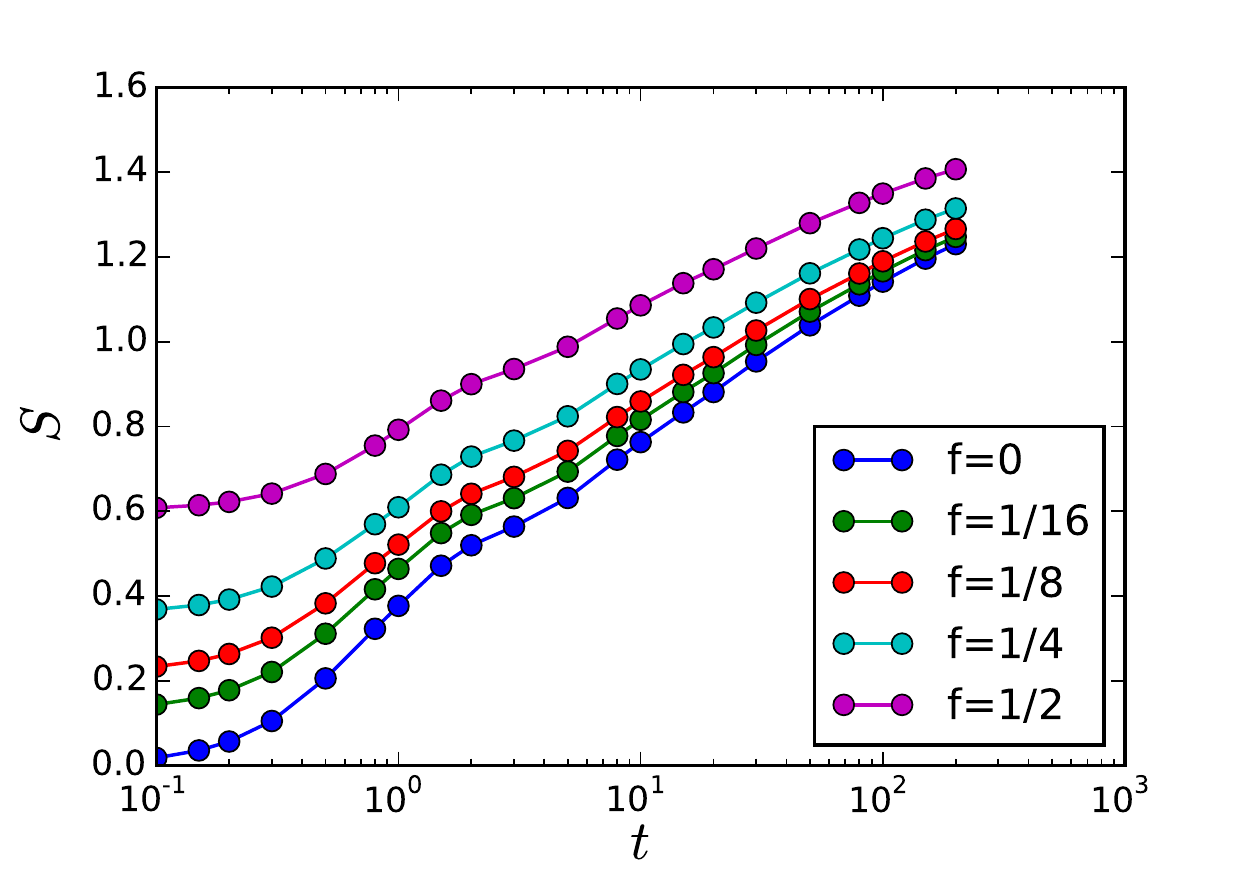}
\caption{Entanglement growth as a function of time for the projected problem. The initial condition is obtained by starting from an unentangled product state and {\it projecting out} an energy window containing a fraction $f$ of the eigenstates, and the time evolution is with respect to an effective Hamiltonian obtained by performing the same projection on (\ref{heisenberg}). Note that the projection increases the entanglement of the initial condition and turns the Hamiltonian into a non-local operator, the time evolution remains logarithmic, suggesting a residual local integrability. Results obtained by exact diagonalization on a  system of $N=14$ spins at $h=4$, with 40 disorder realizations. }
\label{Stime}
\end{figure}
In the FMBL phase entanglement grows logarithmically with time \cite{bardarson}- a numerical observation that was a valuable clue to the emergent integrability of the FMBL phase, since such logarithmic growth naturally arises from dephasing interactions between exponentially localized LIOMs \cite{lbits, Serbynlbits}. We have explored the growth of entanglement starting from unentangled initial conditions,  but in a modified problem where the initial condition and the Hamiltonian generating time evolution have been {\it projected} onto an energy window containing a fraction $(1-f)$ of the eigenstates (Fig.\ref{Stime}).   We find that the projection increases the entanglement entropy of the initial state. However, the {\it dynamics} of the entanglement entropy is unchanged, and we continue to observe a logarithmic growth of entanglement. Given that projection turns the effective Hamiltonian into a non-local operator, this is a surprising result. It leads us to conjecture that the {\it effective} (projected) Hamiltonian continues to show an emergent local integrability i.e. that this notion is well defined even when one only has access to a set of localized states that is a proper subset of the full energy spectrum.

%%%%%%%%%%%%%%%%%%%%%%%%%%%%%%%%%%%%%%%%%%%%%%%%%%%%%%%%%5
\section{Constructing local integrals of motion}
\label{sec::Chandran}
Once we project out a portion of the Hilbert space, we can no longer construct strictly local operators. This is easiest to see by considering the limit $f \rightarrow 1$ when we retain only a single eigenstate. In this limit the only operator we can construct is the (completely nonlocal) projector on that eigenstate. We are interested primarily in the opposite limit $f \rightarrow 0$; however strictly local operators are still unavailable, because a strictly local operator must act as the identity away from a compact region of real space, and the resolution of the identity involves {\it all} the eigenstates of the Hamiltonian (some of which have been projected out). 

To make progress, we introduce the notion of a {\it quasi-local} operator. To make this notion precise, note that any operator $O$ can be decomposed as 
\begin{equation}
O\equiv O^{A}+O^\bot, \label{decomp}
\end{equation}
where $O^A$ has support only on some contiguous subsystem $A$ containing $N_A$ sites. An operator is local if $\| O^\bot \|\rightarrow 0$ exponentially as $N_A$ is increased, where $\|...\|$ denotes the Frobenius norm. In contrast, we define an operator to be `quasi-local' if
\begin{equation}
\lambda \equiv \frac {\| O^A \|^2}{\| O \|^2} 
\label{lambdadef} 
\end{equation}
is non-zero, as $N\rightarrow\infty$ (while keeping $N_A$ fixed), since in this case the operator has a non-zero {\it weight} in a compact region of real space. 

It is useful to note that given a set of quasi-local operators the `local weight' $\lambda$ can be brought closer to one by adapting a method from Ref.~\cite{Prelovsek}. In this method, we take a set of quasi-local operators $O_{i}$  where the $i$ lie in a region $A$ and construct the matrix $K_{ij}\equiv\langle O_{i}|O_{j}\rangle$. The eigenvectors of this matrix give us a new orthogonal set of operators, whose eigenvalues are the $\lambda$ for these operators. We then take the operator with the largest $\lambda$ as an effective quasi-local operator where the local part has support on a region of size $N_A$. Using larger sets of operators increases $\lambda$, at the cost of expanding the region where the local part has support, and also increasing the computational cost of the algorithm. 

\begin{figure} 
\includegraphics[width=\linewidth]{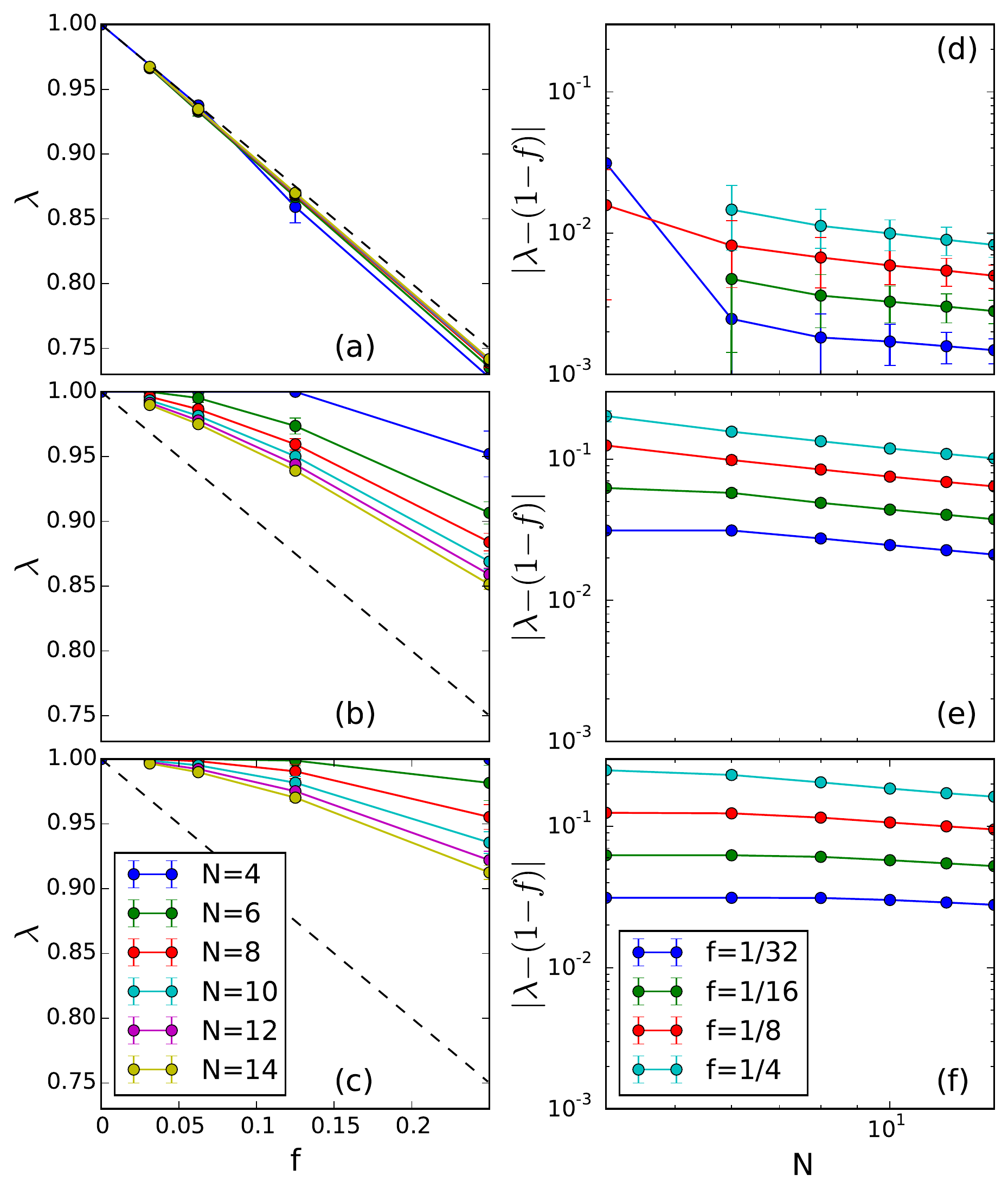}
\caption{Quasi-local operators in a projected Hilbert space, not constrained to commute with the Hamiltonian. Figs.(a-c) shows the local weight $\lambda$ as a function of $f$ (averaged over position) for a sequence of system sizes $N$, and with $N_A = 1 (a), 2(b), 3(c)$. Increasing $N_A$ brings $\lambda$ closer to one, but the `improvement' grows smaller as system size is increased.  Figs. (d-f) shows $|\lambda-(1-f)|$ as a function of system size $N$ with $N_A = 1 (d), 2(e), 3(f)$. When $N$ is made large at fixed $N_A$ it appears that $\lambda \rightarrow 1-f$. %The fact that the points following a straight line (with negative slope) on these double logarithmic plots implies that the quantities are decaying to zero. 
Data was taken for $h=8$ (though the results are largely independent of $h$) and 200 disorder realizations. }
\label{lambda}
\end{figure}

We therefore proceed as follows: we start with set of local operators and `delete' the portion that has any matrix element to a `projected out' energy window containing a fraction $f$ of the eigenstates. Fig. ~\ref{lambda}(a) shows the $\lambda$ values obtained from this procedure, projecting out a fraction $f$ of the states at the top of the spectrum. 
Significantly increased $\lambda$ values are obtained by employing the above-mentioned trick from Ref.~\onlinecite{Prelovsek} for $N_A>2$ [Fig.~\ref{lambda}(b)] and further increased for $N_A=3$ [Fig.~\ref{lambda}(c)]. 

% Fig \ref{lambda}(b,c) show the increased $\lambda$ values obtained by employing the above-mentioned trick from Ref.\cite{Prelovsek}. Results were obtained for $N_A=1,2,3$, and The sets of operators input to the algorithm included every non-identity operator which conserves the total $S^z$ in $A$ (when $N_A=1$ then $O_i=\sigma^z_i$).
%
 It can seen in Fig.~\ref{lambda}(d-f) that $\lambda$ is non-zero even as $N$ grows large at fixed $N_A$ (i.e. quasi-local operators exist after projecting out a portion of the Hilbert space).
 Though at large $N_A/N$ $\lambda \sim 1-f$, by increasing $N_A$ we can make $\lambda$ even larger at the cost of increasing the computational cost and the size of the region where the local part has support.
 Note that at present these are simply quasi-local operators - not constants of motion. 

We now take our set of quasi-local projected operators from the method above and turn them into integrals of motion using the method from Ref.~\cite{Chandran}. The centerpiece of this method is the observation that any operator $O$ can be made to commute with a Hamiltonian by performing the Heisenberg time evolution (in our case with respect to the \emph{projected Hamiltonian}) and averaging over time.  This yields an integral of motion $\tilde{O}$ defined as 
\begin{equation}
\tilde{O}\equiv \sum_E^{~~~~\prime} |E\rangle\langle E|O |E\rangle\langle E|,
\label{timeavg}
\end{equation}
where $E$ is an energy eigenstate of the system, and the prime on the sum is because we are summing over only a fraction $(1-f)$ of the eigentstates. 
With time averaging, the off diagonal matrix elements average to zero, and only diagonal elements survive.
An operator that is diagonal in the energy eigenbasis trivially commutes with the Hamiltonian. In this way, one can always turn a set of (quasi)local operators into integrals of motion. The key question is: does this time average produce a quasi-local integral of motion, or something completely non-local? 

Fine-grained information about the structure of the commuting operators produced by this procedure is contained in the two point correlator between an operator and its time-averaged version:
\begin{equation}
M_{ij}\equiv 2^{-N}Tr(\tilde {O}_i O_j)
\end{equation}
which may be interpreted as acting on the system with the operator $O_i$, then evolving in time and finally measuring $O_j$. In the MBL phase $O_j$ will not be affected by the application of $O_i$ when $i$ and $j$ are far apart, and therefore we expect $M_{ij}$ to decay exponentially in $|i-j|$. This need not be the case in the thermal phase. % In contrast, in the thermal phase $M_ij$ will be independent of distance, as the effects of the perturbation will propagate through the entire system. 

\begin{figure}
\includegraphics[width=\linewidth]{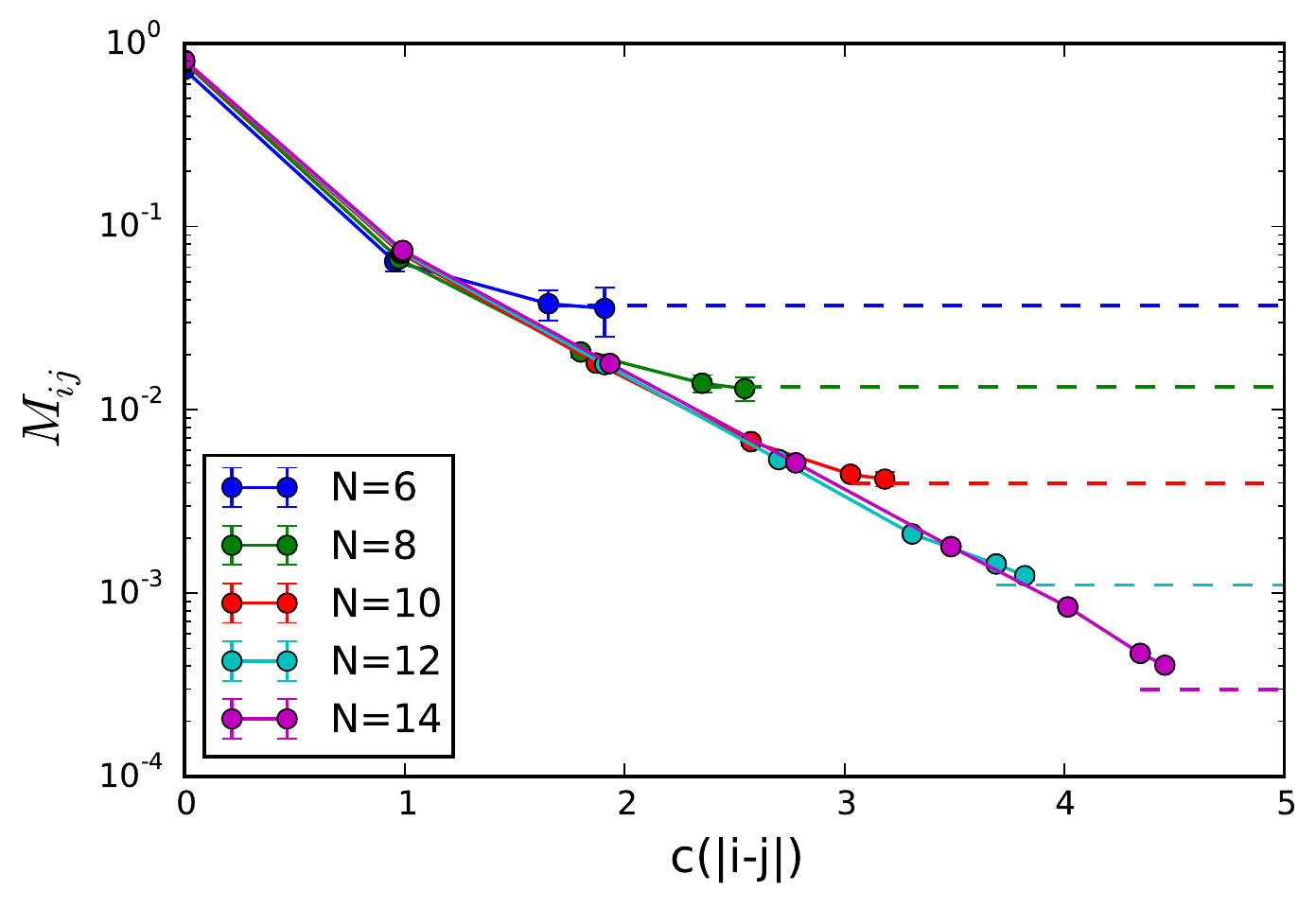}
\caption{Solid lines show the weight of the integral of motion $\tilde O$ as a function of distance for the case where $\hil_{trunc}$ does not grow with system size $N$, therefore $f \sim exp(-N)$. 
The $y$-axis of this plot is on a log scale, and the $x$-axis is `chord distance' (see text), straight lines on such a plot indicate exponential decay, and this is what we observe.
Dashed lines show the corresponding weight for the parent quasi-local operator $O$, which does not commute with the Hamiltonian.
Data was taken for $N_A=1$, $h=8$, $N_{trunc}=7$ and $200$ disorder realizations. 
Note that $M_{ij}$ decays exponentially at small distances before saturating to a constant background (indicated by the dashed lines), but this background itself is exponentially small in $N$. 
In the thermodynamic limit the background disappears, such that the integral of motion $\tilde O$ is truly an LIOM, with exponential tails out to long distances i.e. when the number of states projected out is held constant as we increase system size, we can construct truly localized integrals of motion. }
\label{smallf}
\end{figure}

We begin by evaluating $M_{ij}$ for integrals of motion obtained as above, for the case where the number of states removed from the many-body spectrum, $\hil_{trunc}$, is constant in system size $N$. This implies $f\sim exp(-N)$, since the total Hilbert space dimension $\hil=2^N$. Fig.~\ref{smallf} shows $M_{ij}$ for a system with $7$ states removed, for $N_A=1$ and several system sizes, deep in the MBL phase. The dashed lines show the background:
\begin{equation}
{\rm background}=\frac{1}{N(N-1)}\sum_{i\neq j} 2^{-N}Tr(O_iO_j), 
\label{background_eq}
\end{equation}
which is the two point correlator for the non-time averaged operators (which does not commute with the Hamiltonian). For unprojected operators, this quantity would be non-zero only when $i=j$, but since our operators are quasi-local there is a global `background' value which is small and independent of $|i-j|$ when $i\neq j$.
%This background is an artifact of the fact that we are not studying local operators, but instead quasi-local operators in the sense of Eq.~\ref{lambdadef}.
%Whether or not local operators exist in this truncated subspace is an open question. 
%A local operator will have an $M_{ij}$ that decays exponentially, but since we started with quasi-local operators $M_{ij}$ will not drop below the background value. 
What we are looking for is an $M_{ij}$ which decays exponentially until it reaches the background value, even when the sum in Eq.~(\ref{timeavg}) does not extend over all states. Such an operator is as local as it can be, given that we are working with an imcomplete Hilbert space. Indeed we see that $M_{ij}$ decays exponentially at short distances before saturating to a constant `background' at large distances, with the background value being set by the quasi-locality of the parent operator. 
The $x$-axis of Fig.~\ref{smallf} is rescaled by the `chord distance', $c(x)=N/\pi \sin(\pi x/N)$. Such a rescaling has been found to lead to straight lines for exponentially decaying quantities plotted on a semi-log plot for one-dimensional systems with periodic boundary conditions.

\begin{figure}
\includegraphics[width=\linewidth]{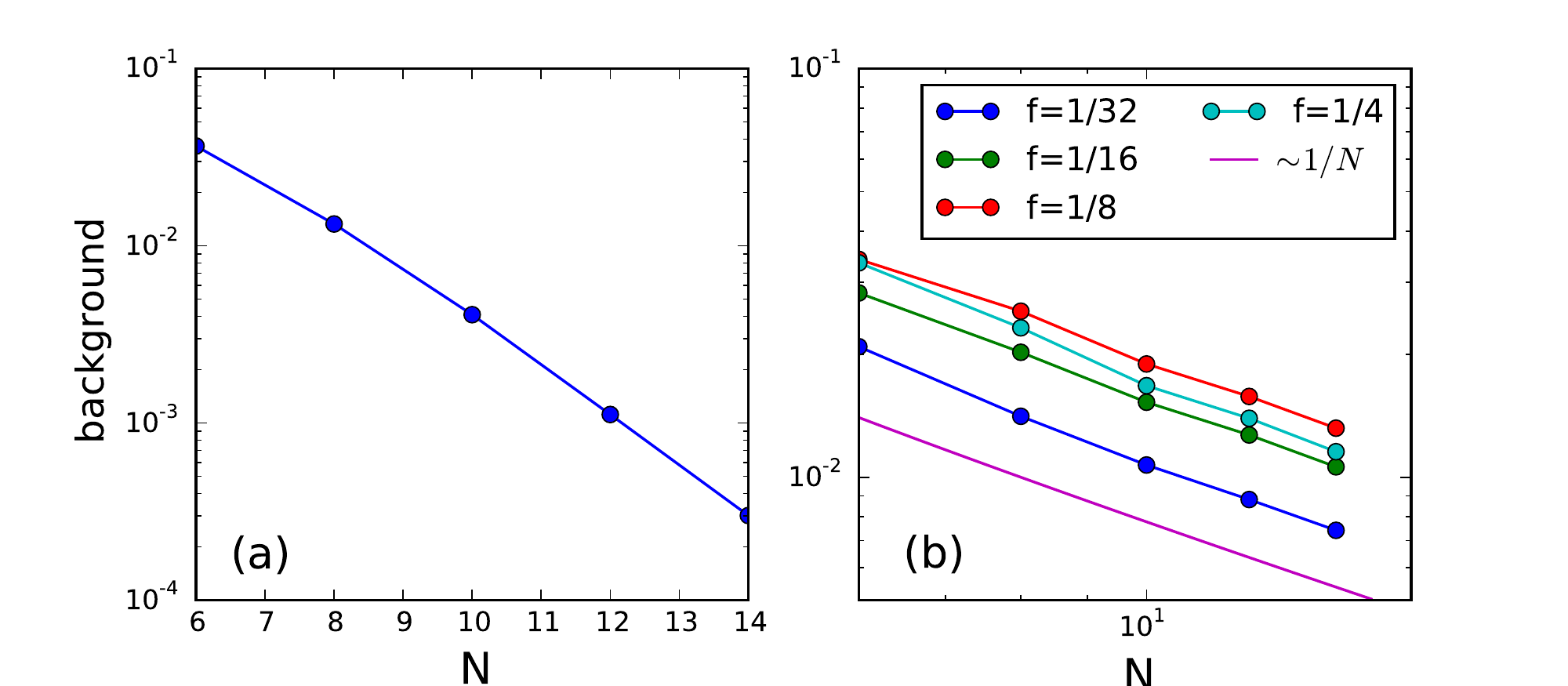}
\caption{The dependence on system size of the global background on a single site, for (a) the case where the number of states projected out is held constant, so $f$ vanishes exponentially in the thermodynamic limit  (semi-log plot) and (b) the case where $f$ is independent of system size (log-log plot). In (a) the background decays exponentially with $N$, while in (b) it decays as $1/N$. Data was taken for $h=8$, $N_A=1$ and $40$ disorder realizations.}
\label{background}
\end{figure}

The size of the `background' in Fig.~\ref{smallf} decays exponentially with $N$, [see Fig. \ref{background}(a)] such that in the thermodynamic limit the above procedure generates true integrals of motion, with exponential tails out to long distances. We expect similar results whenever $f\rightarrow 0$ in the thermodynamic limit, i.e. projecting out a number of many body states that grows more slowly with system size $N$ than $d^N$ (where $d$ is the local Hilbert space dimension) should not interfere with {\it exact} emergent local integrability.

\begin{figure} 
\includegraphics[width=\linewidth]{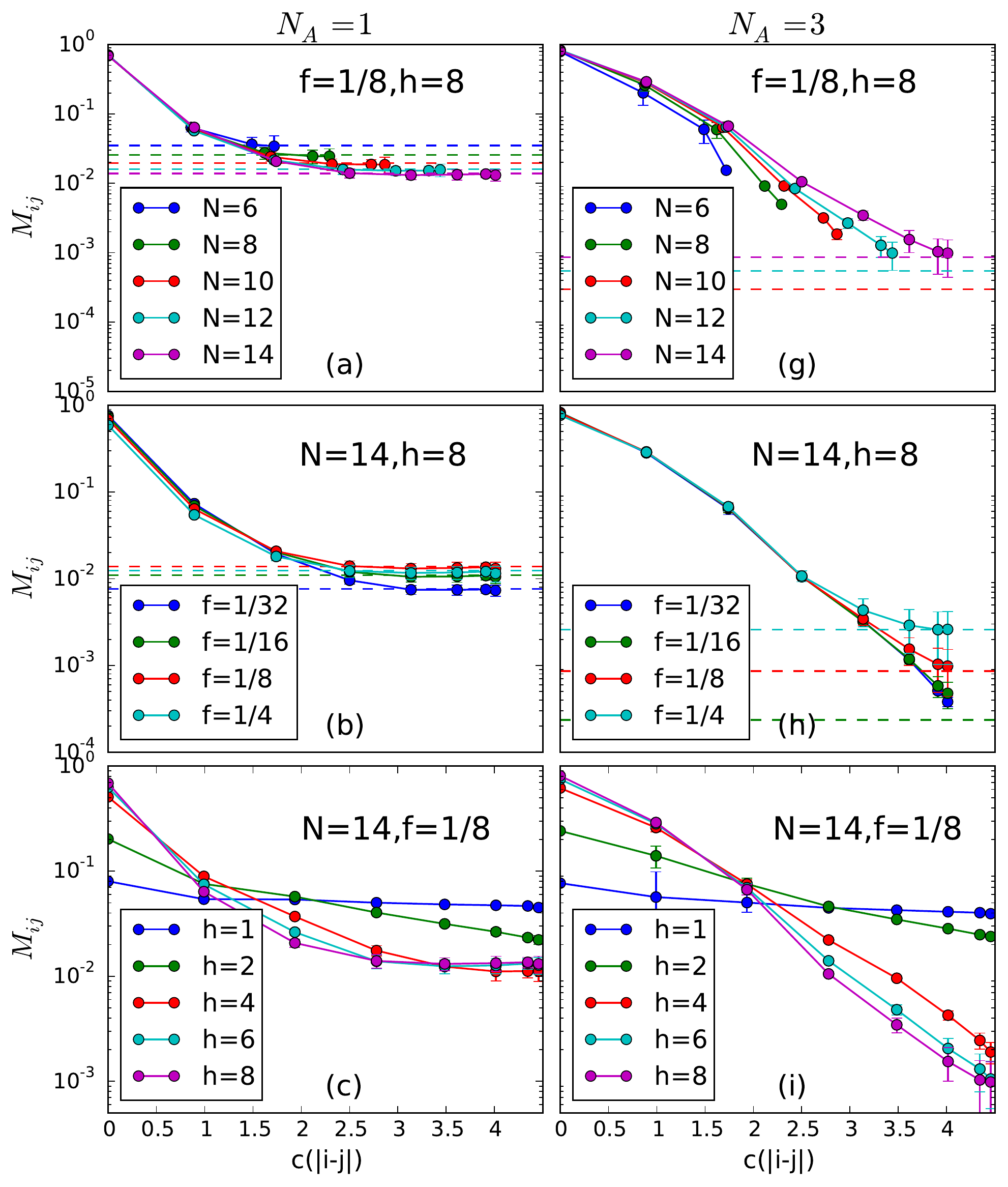}
\caption{ $M_{ij}$ as a function of $c(|i-j|)$ (see text) for different sets of operators, system sizes, and projected fractions $f$. Solid lines show $M_{ij}$ for time averaged operators $\tilde O$ (which commute with the Hamiltonian) and dashed lines show the corresponding overlaps for the parent projected operators $O$ (which do not commute with the Hamiltonian). 
The columns of Fig.~\ref{Mij} correspond to $N_A = 1$ and $3$. 
 }
\label{Mij}
\end{figure}
In Fig.~\ref{Mij} we show the $M_{ij}$ obtained when the {\it fraction} $f$ of states projected out is held constant as the system size is made large. 
The first row shows data analogous to Fig.~\ref{smallf}, but with $f$ held constant as $N$ is varied. We can see that the weight of the integral of motion $\tilde O$ originally decays exponentially, before saturating to a background set by the global dressing of projected (non-time averaged) operators (dashed lines). This is significantly easier to see with larger $N_A$ (with $N_A=1$ the background is so large as to essentially obscure the signal). The second row fixes system size $N=14$ and instead varies the projected fraction $f$. The short distance exponential decay of the quasi-LIOM is observed to be independent of $f$, but the `global background' is controlled by $f$, and is set simply by the global dressing of the projected non-time averaged operator (which scales as $\sim f$ at small $f$, see Fig.~\ref{lambda}). The final row shows data with fixed $N=14$ and $f=1/8$ but varies the disorder strength $h$. Exponential decay of the overlaps, corresponding to (quasi)LIOMs, is observed only for strong enough disorder. For weak disorder $h=1$, when the set of eigenstates we work with includes delocalized states, there is no exponential decay of the overlaps i.e. the integrals of motion obtained from the time averaging procedure are fully global quantities.

 The main message of Fig.~\ref{Mij} is that as long as the states we have access to are all localized, the time averaging to produce integrals of motion does not introduce any \emph{additional} non-locality besides that already introduced by the projection onto a portion of Hilbert space. In contrast, at weak disorder the time averaging produces completely non-local integrals of motion.

We note that when the thermodynamic limit is taken at fixed $N_A$ and $f$, then the weight of all operators outside of $N_A$ is independent of both $N_A$ and $N$ (Fig.~\ref{lambda}), so the level of the `background' on \emph{ each site} should be $\approx 1/N$, as can be seen in Fig.~\ref{background}(b). Therefore if we could access very large sizes, the background would less of a problem. The use of $N_A>1$ in this section can be thought of as a trick which lowers the background at small sizes, allowing the exponential decay of $M_{ij}$ to be observed. 

Thus, we conclude that a set of localized states which is a proper subset of the energy spectrum is sufficient for emergent local integrability, at least in a `Fermi liquid' like sense where the integrals of motion have non-zero weight in a compact region of space. In particular, when one starts with an FMBL system and projects out a vanishing {\it fraction} of the states (i.e. when the number of states projected out grows more slowly than $d^N$ in the thermodynamic limit $N\rightarrow \infty$), one can still construct using the remaining localized states an extensive number of truly local integrals of motion. On the other hand, when the fraction of states projected out is held constant as the thermodynamic limit is taken then one obtains quasi-LIOMs whose weight decays exponentially as a function of distance at short distances, before saturating to a non-zero background value, with the total weight in the background being $\sim f$. 
%This is illustrated by Fig\ref{lambda2}, which is the analog of Fig.\ref{lambda} for time averaged operators.

Alternative algorithms for constructing LIOMs also exist, such as the celebrated algorithm of Pekker and Clarke for constructing lbits (LIOMs which are related to the original degrees of freedom by a local unitary transformation), see Ref.\cite{PekkerClarke}. We have attempted this method for our projected problem. A crucial step in the algorithm involves matching the eigenstates to configurations of lbits. Since we have an incomplete set of eigenstates this matching induces a non-local component to each lbit (making them quasi local bits). We have found that this non-local component is large enough to prevent us from doing similar analysis to that done above for quasi local integrals of motion. 

%The results obtained from this algorithm are discussed in Appendix\ref{lbits}, but this algorithm appears less successful for the projected problem than the time averaging procedure discussed above.

%%%%%%%%%%%%%%%%%%%%%%%%%%%%%%%%%%%%%%%%%%%%%%%%%%%%%%%%%5
\section{Discussion}

We now discuss some implications of our results. One implication is for the longstanding problem of whether in the strong disorder regime truly {\it all} the eigenstates of Eq.\ref{heisenberg} are localized or only {\it almost all}, with a vanishing fraction of the states in the thermodynamic limit being delocalized (see e.g. the discussion in \cite{Bauer}). In light of our results, we can say that this distinction is unimportant for the dynamics, since a vanishing fraction of non-localized states can always be projected out, and one can still construct an extensive number of strictly local integrals of motion from the remaining localized states \footnote{The situation where projection changes the topological properties of the problem is more subtle, and is beyond the scope of the present work.}. 

Our results also apply directly to some settings where projected Hilbert spaces naturally arise. One example is situations where one is interested only in a low energy subspace e.g. studies of the strong coupling limit of the disordered Hubbard model where one forbids double occupancy (Gutzwiller projection).  Our results imply that in the low energy subspaces obtained thereby one can still have emergent local integrability, at least in the sense of an extensive number of quasi-LIOMs 
\footnote{Strictly our results only apply when the low energy subspace contains a finite fraction of the total states, which will be true only for models at sufficiently low density}. As such, they may have applications to strongly correlated materials. 

%\addSG{ I deleted a paragraph here since it is pretty much the same as one in the intro}
%Our results may also have relevance for gauge theories, where the imposition of a local gauge constraint  `projects out' some states. Additionally, they reopen questions addressed by recent no go theorems that demonstrate the incompatibility of non-Abelian symmetries and certain types of order \cite{pottervasseur, pottervishwanath} with strict LIOMs. Can these symmetries or types of order be compatible with a weaker form of emergent integrability involving quasi-LIOMs, in the sense we have discussed? 

Our results also have implications for systems where a finite fraction of the spectrum is delocalized, and separated from the localized states by many body mobility edges. Such mobility edges generically arise in perturbative and numerical analyses at intermediate disorder (see however the arguments of Ref.\cite{bubbles}, which are discussed at length in Appendix \ref{bubbles}). However, it is at present unclear whether the notion of emergent integrability, so fundamental to our understanding of the FMBL regime, applies when many body mobility edges are present. An application of our results implies that a version of emergent integrability should continue to hold in this regime, since one could simply project out the delocalized states and obtain quasi-LIOM's from the remaining localized states. The quasi-LIOM's would however have a `global dressing' that would contain at least a fraction $f$ of the weight of the operator, where $f$ is lower bounded by the fraction of states that is delocalized. Thus, as more and more of the spectrum becomes delocalized, the quasi-LIOM's obtained from our procedure have ever smaller `local residue,' and cease to be local in any sense when almost all of the states are delocalized.  

A more careful discussion of the implications of our result for the problem with mobility edges requires a discussion of the minimal requirements for emergent integrability in the presence of mobility edges. The original l-bit ansatz \cite{lbits, Serbynlbits} holds that in the FMBL phase, when all eigenstates are localized, the Hamiltonian can be rewritten in terms of operators $\tau^z_i$ and takes the form
\begin{equation}
H = \tilde h_i \tau^z_i + K_{ij} \tau^z_i \tau^z_j + \Lambda_{ijk} \tau^z_i \tau^z_j \tau^z_k + ... \label{l-bitHamiltonian}
\end{equation}
where repeated indices are summed over. The operators $\tau^z$ are related to the physical bits (pbits) $\sigma$ by a local unitary transformation with exponential tails, and the coefficients $(K_{ij}, \Lambda_{ijk}, ...)$ fall off exponentially with distance between the furthest apart operators. Further, the Hamiltonian contains only $\tau^z$ operators, such that every $\tau^z_i$ commutes with the Hamiltonian and is localized in real space. Clearly in a model with mobility edges, the Hamiltonian cannot be rewritten in the form Eq.\ref{l-bitHamiltonian} because an l-bit Hamiltonian of this form corresponds to localized dynamics at {\it every} energy, whereas the existence of mobility edges implies that the dynamics is localized in certain energy windows, but delocalized in other energy windows. What might this mean at the operator level? One possibility is that there might exist an extensive number of localized operators $\hat O_i$ which do not necessarily commute with the Hamiltonian, but which satisfy 
\begin{equation}
\hat P_l [\hat O_i, H] \hat P_l = 0 \label{minimal1}
\end{equation}
where $\hat P_l$ is the projector onto the space of localized states. This implies an extensive number of local integrals of motion when the system is prepared in the localized subspace, and is sufficient for local integrability in the regime with mobility edges. Our work, however, imposed a stronger condition, asking for (quasi)local operators $\hat O_i$ which {\it act only within the localized subspace} and which additionally commute with the Hamiltonian. Can we get more local integrals of motion if we relax our condition to (\ref{minimal1}), allowing the operator to act within the space that we had projected out and demanding only that the projected commutator with the Hamiltonian vanishes? 

A general operator $\hat O$ may be decomposed as $\hat O = O_{A} \otimes 1 + 1 \otimes O_B + O_{AB}$, where $O_A$ has non-trivial matrix elements only within energy window $A$, $O_B$ has non-trivial matrix elements only within energy window $B$, and $O_{AB}$ has non-trivial matrix elements between the two energy windows. Since the Hamiltonian is diagonal in the energy basis, (\ref{minimal1}) essentially amounts to the condition $[O_A, H] = 0$ i.e. there are no constraints on $O_B$ and $O_{AB}$. %If $\hat O$ is a constant of motion under time evolution by $H_A$, then $O_{AB} = 0$, since off diagonal matrix elements in energy space will oscillate at a frequency set by the energy difference. Thus, an integral of motion can be decomposed simply as $\hat O = O_{A} \otimes 1 + 1 \otimes O_B$. 
Our procedure sets $O_B$ and $O_{AB}$ to zero, and finds $O_A$ by time averaging. This procedure yields a quasi-LIOM with a global dressing. Could this global dressing be reduced or even eliminated by `adding in' a suitable $O_B$ and $O_{AB}$, which are {\it not} constrained to commute with the Hamiltonian, but are simply chosen to make the integral of motion as local as possible? 

 In this work, the original Hamiltonian was fully MBL, and therefore we can certainly find completely localized operators satisfying (\ref{minimal1}) - these are simply the lbits which commute with the Hamiltonian even before projection. If the energy window $B$ contained delocalized states, we could instead perform some kind of optimization to find the best $O_{B}$ and $O_{AB}$. If the region of Hilbert space spanned by the energy window $B$ were the same as that spanned by the same energy window in the FMBL case, then one can find the {\it same} $O_B$ and $O_{AB}$, leading to perfectly local lbits (which commute with the Hamiltonian only after projection on energy window $A$). %However, this clearly cannot happen, since perfectly local operators that commute with the Hamiltonian (without projection) implies localization at all energy densities, and we just assumed that energy window $B$ contained delocalized states. 
However when the states in energy window $B$ go through a localization-delocalization transition, the region of Hilbert space spanned by these states may change (this certainly happens when $B$ contains a single eigenstate), so that we may no longer be able to precisely cancel the `global background.' How good a job one can do of canceling the global background in systems with mobility edges remains an interesting open problem. Another subtlety that must be noted is that if the region of Hilbert space spanned by $B$ changes, so does the region of Hilbert space spanned by $A$ i.e. the existence of delocalized states somewhere in the spectrum changes at least to some extent the structure of localized states elsewhere in the spectrum \footnote{Indeed a particularly extreme scenario \cite{bubbles} holds that if delocalization occurs anywhere in the spectrum, it occurs everywhere }. While we believe this effect should be modest at least when $f$ is small, to what extent it complicates application of our results to the problem with mobility edges is an interesting question for future work \footnote{Numerically the situation is murky since the many body mobility edge in the small system sizes that can be readily accessed is not very sharp}. %and fresh thinking will be required to conclusively settle this issue.
 %A similar approach could also be attempted in the algorithm of Ref.~\onlinecite{PekkerClarke}.

\label{discussion}

{\bf Acknowledgements:}
We acknowledge conversations with Anushya Chandran, Chetan Nayak, David Pekker and Rahul Roy. We thank Markus Muller for feedback on the manuscript. SG and RNB acknowledge support from Department of Energy  BES Grant DE-SC0002140.
RNB thanks the Aspen Center for Physics for hospitality during the writing of this manuscript.

\appendix
\section{Quasi Local Integrals of Motion when the middle of the spectrum is projected out}
\label{middle}
\begin{figure} 
\includegraphics[width=\linewidth]{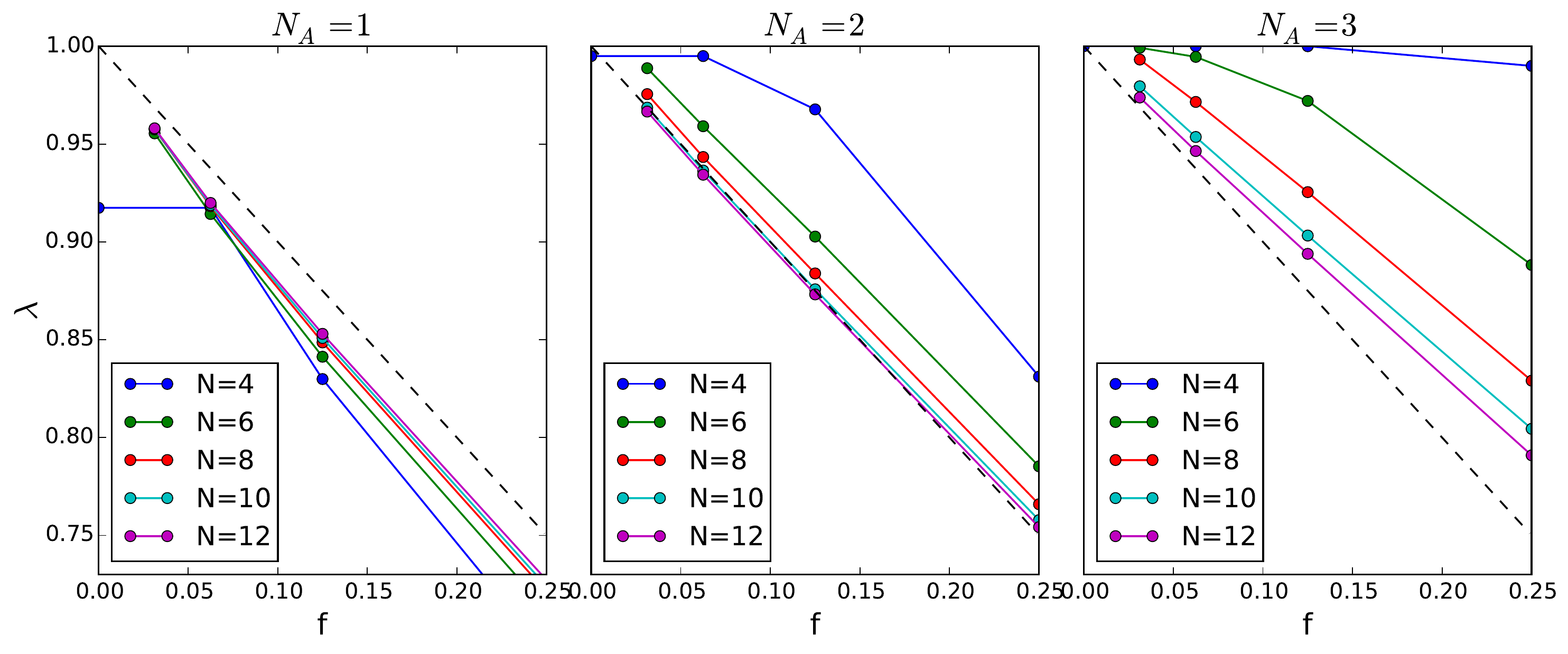}
\caption{The equivalent of Fig.~\ref{lambda}, but for a system where the middle states have been projected out instead of states at the upper end of the energy spectrum. The qualitative physics is the same: increasing $N_A$ increases $\lambda$ at finite sizes, with this effect vanishing the thermodynamic limit. But the size of this effect is is much smaller, which makes it harder to extract the relevant physics from the background. }
\label{lambda_middle}
\end{figure}

\begin{figure} 
\includegraphics[width=\linewidth]{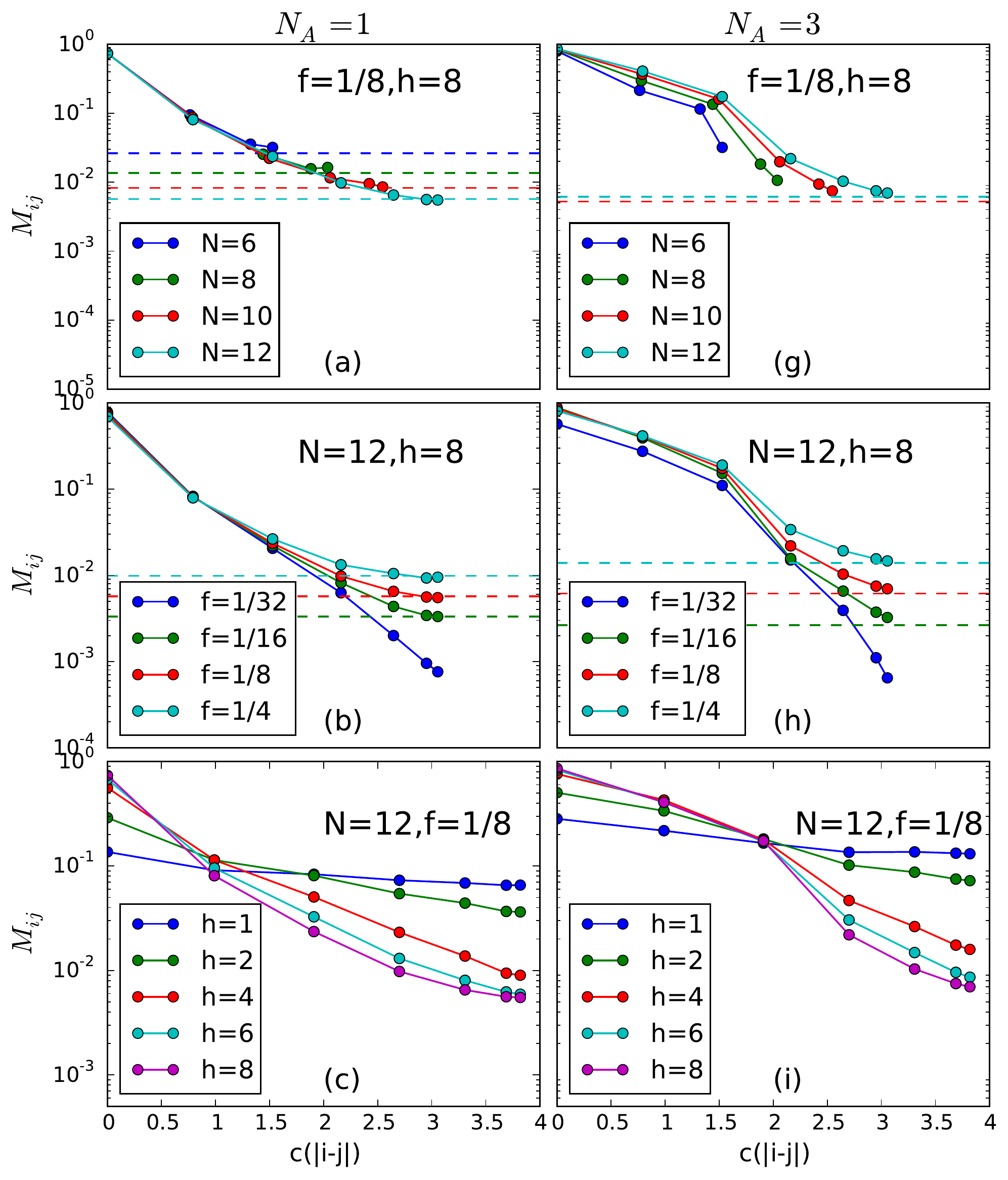}
\caption{The equivalent of Fig.~\ref{Mij}, but for a system where the middle states have been projected out instead of states at the upper end of the energy spectrum. At finite sizes this choice raises the background, making the exponential decay more difficult to see. }
\label{M_middle}
\end{figure}

In the main text we chose to project out the highest energy eigenstates, as opposed to states in the middle of the band, which would be better motivated in the context of mobility edges. This is mainly because projecting out the `top states' leads to less severe finite size effects. Since we are starting from a fully many-body localized regime, it should not matter which states we project out. However, at the finite system sizes that we study we devoted a fair amount of effort to reducing the non-local `background' that came from projecting into only a finite fraction of the Hilbert space. To reduce this background, we want to maximize $\lambda$, which is the overlap between a projected operator, and all the unprojected operators in a region $N_A$. In the thermodynamic limit $\lambda$ is equal to $1-f$, the fraction of states remaining in the Hilbert space. But at small sizes, we found that $\lambda$ was significantly enhanced for $N_A>1$ (see Fig.~\ref{lambda}). The enhancement is a non-universal, finite-size effect, which we find very useful. In Fig.~\ref{lambda_middle}, we show the equivalent of Fig.~\ref{lambda} for the situation where the middle of the energy spectrum is projected out, instead of the end. We see that $\lambda$ is reduced at $N_A>1$. This in turn raises the background, as can be seen in Fig.~\ref{M_middle}. This larger background makes the exponential decay of $M_{ij}$ harder to see. Therefore, although we don't believe that the choice of energy window to project out changes any of the physics, in the main text we have projected out a window of states at the `top' of the many body spectrum. 

\section{On the existence of mobility edges}
\label{bubbles}
It has recently been claimed \cite{bubbles} that many body mobility edges may not exist. It is argued in \cite{bubbles} that the perturbative calculations showing mobility edges \cite{BAA, Gornyi} miss certain  effects that destroy mobility edges, and the numerical investigations showing mobility edges all miss these self-same effects due to finite size limitations. Instead it is claimed that localization is an `all or nothing' phenomenon in the thermodynamic limit - either all states are localized, or all states are delocalized. However, the arguments in \cite{bubbles}, while fascinating and provocative, in our view leave open several loopholes (discussed below) which allow for the existence of many body mobility edges.

Let us begin by reviewing the argument from Ref.\cite{bubbles}. It is argued that the existence of a many body mobility edge implies an obstruction to the construction of a perturbative locator expansion for the eigenstates. This is because rare regions (aka bubbles) where a perturbative locator expansion undergoes local resonant breakdown because the local energy density is on the delocalized side of the mobility edge, can move around at high orders in perturbation theory, such that the eigenstates necessarily contain a percolating network of resonances. This then is argued to be inconsistent with localization. 

This argument can be critiqued on two levels. Firstly, the percolation of resonances implies breakdown of perturbation theory, but does not imply delocalization of the eigenstates. For example, a non-interacting localization problem on a two dimensional lattice with binary disorder has percolating networks of resonances that formally `break' a naive locator expansion, but the eigenstates are nonetheless still localized. (we would like to thank John Imbrie for providing this example). This possibility was considered in \cite{bubbles} and was asserted to not be a real problem, but in our view this is a potential loophole in the argument. Secondly, even if the {\it eigenstates} are delocalized, this does not necessarily mean that the {\it dynamics} are delocalized \cite{NGH, JNB, lstarbits}.  Indeed localization in the dynamics is far more robust than localization of eigenstates, as has been discussed in Ref.\cite{NGH, JNB} and most recently and eloquently in \cite{lstarbits}. The argument for delocalized dynamics is founded on a mental model whereby the bubbles transport energy, violating the central prediction that the localized phase should be a thermal insulator. However, the leap to such a `transport' picture assumes that the `bubbles,' which were regions where the local energy density was on the delocalized side of the mobility edge, could be treated as semiclassical objects. This may be an oversimplifiication given that these `bubbles' are embedded in a localized environment, which does {\it not} decohere the bubbles, which would appear to be a necessary condition for such a semiclassical picture to obtain. %Thus, even if the {\it eigenstates} contain percolating networks of resonances (as Ref.\cite{bubbles} showed) the {\it dynamics} could still be localized.

For all of the above reasons, we believe the arguments in Ref.\cite{bubbles}, while fascinating and provocative, do not rule out mobility edges. We therefore take the existence of many body mobility edges as a working hypothesis, supported by a vast array of perturbative and numerical results. It of course remains possible that the loopholes identified above could be closed. Even in that scenario however, apparent mobility edges would continue to arise in finite size systems, for which our analysis would remain important. 

\section{Additional Numerical Data}
In this section we provide some additional numerical results which further support some of the conclusions drawn in the main text. 

\begin{figure} 
\includegraphics[width=\linewidth]{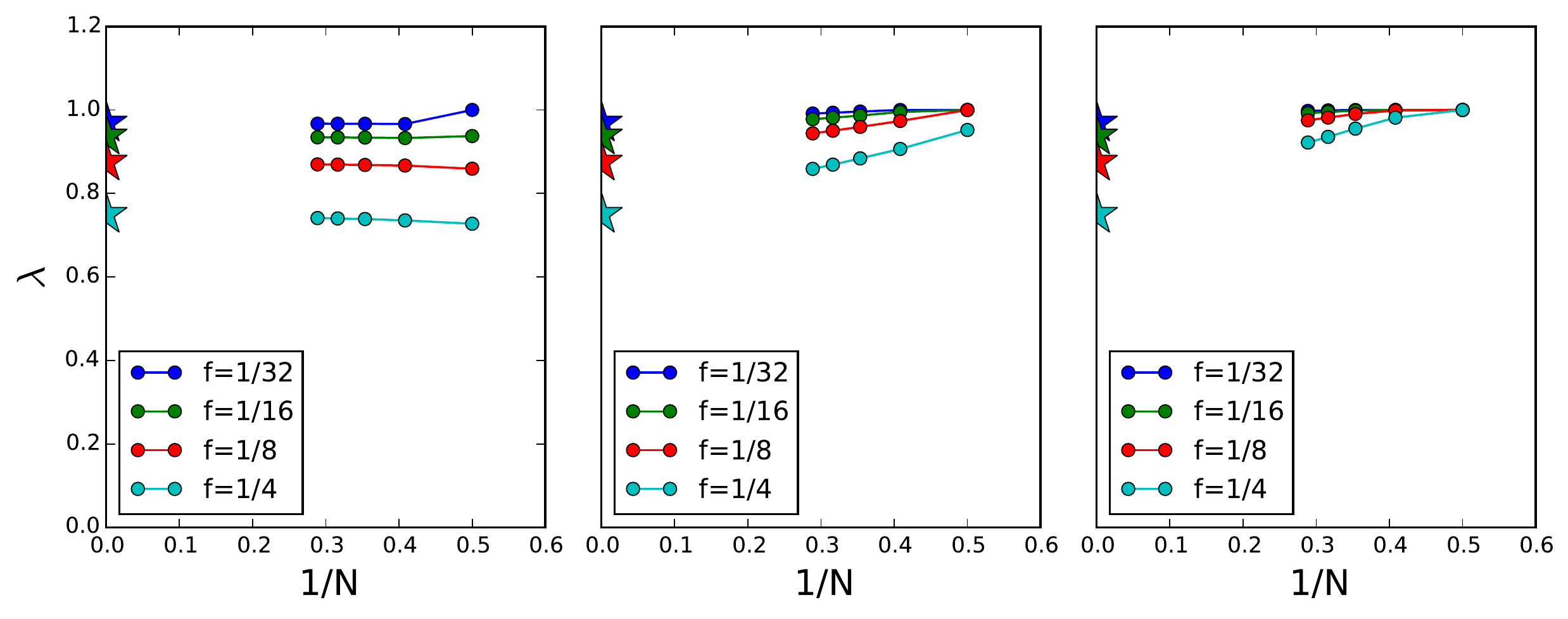}
\caption{The same data as the bottow row in Fig.~\ref{lambda}, but plotted against $1/N$. We see that $\lambda$ does not decay to zero in the thermodynamic limit, and in fact seems to be consistent with the value $(1-f)$ in the thermodynamic limit, indicated by the stars on the y axis}
\label{lambda_vs_1overN}
\end{figure}

In Fig.~\ref{lambda_vs_1overN} we show data equivalent to the bottom row of Fig.~\ref{lambda}, but plotted in a different way: we plot $1/N$ vs. $\lambda$ on linear axes. The purpose of this plot is to further justify that in the thermodynamic limit $\lambda$ is \emph{finite}, and therefore our operators are quasi-local as defined in the main text.

\begin{figure} 
\includegraphics[width=\linewidth]{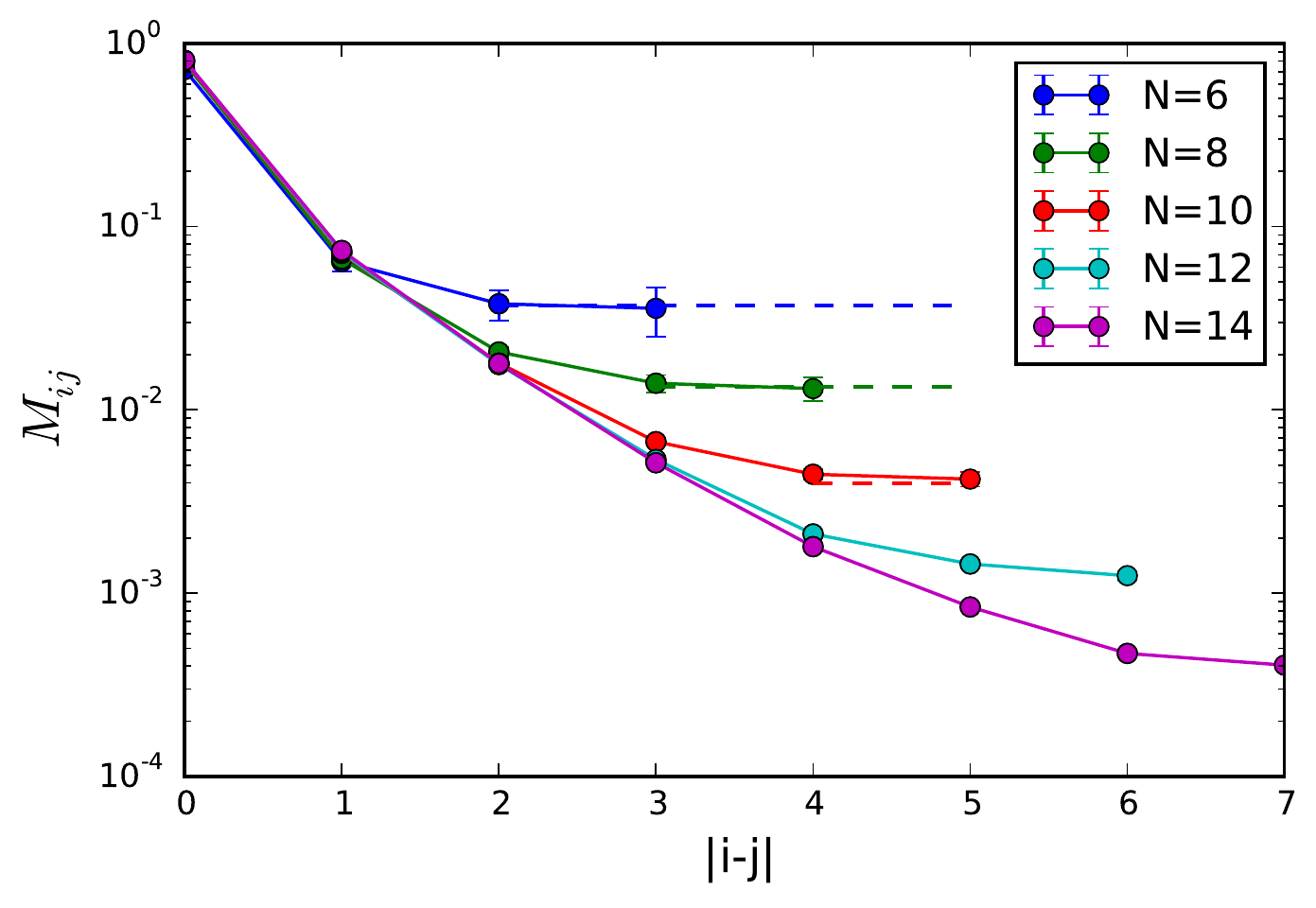}
\caption{ The same data as Fig.~\ref{smallf} but plotted with respect to actual distance (not chord distance). The exponential decay is harder to see but still present, showing that it is not an artifact of our rescaling of the $x$-axis by the chord distance.}
\label{rescaled}
\end{figure}

In Fig.~\ref{rescaled} we show the data in Fig.~\ref{smallf}, but we do not rescale the x-axis by the `chord distance' This results in a flattening of the curves at large $x$ because of the periodic boundary conditions. Though this makes the exponential decay more difficult to see, in the range $2-6$ we still see approximately a straight line (similar to the results in Ref.~\onlinecite{Chandran}) and therefore we conclude that the exponential decay is not an artifact of the `chord distance' rescaling.

\begin{figure} 
\includegraphics[width=\linewidth]{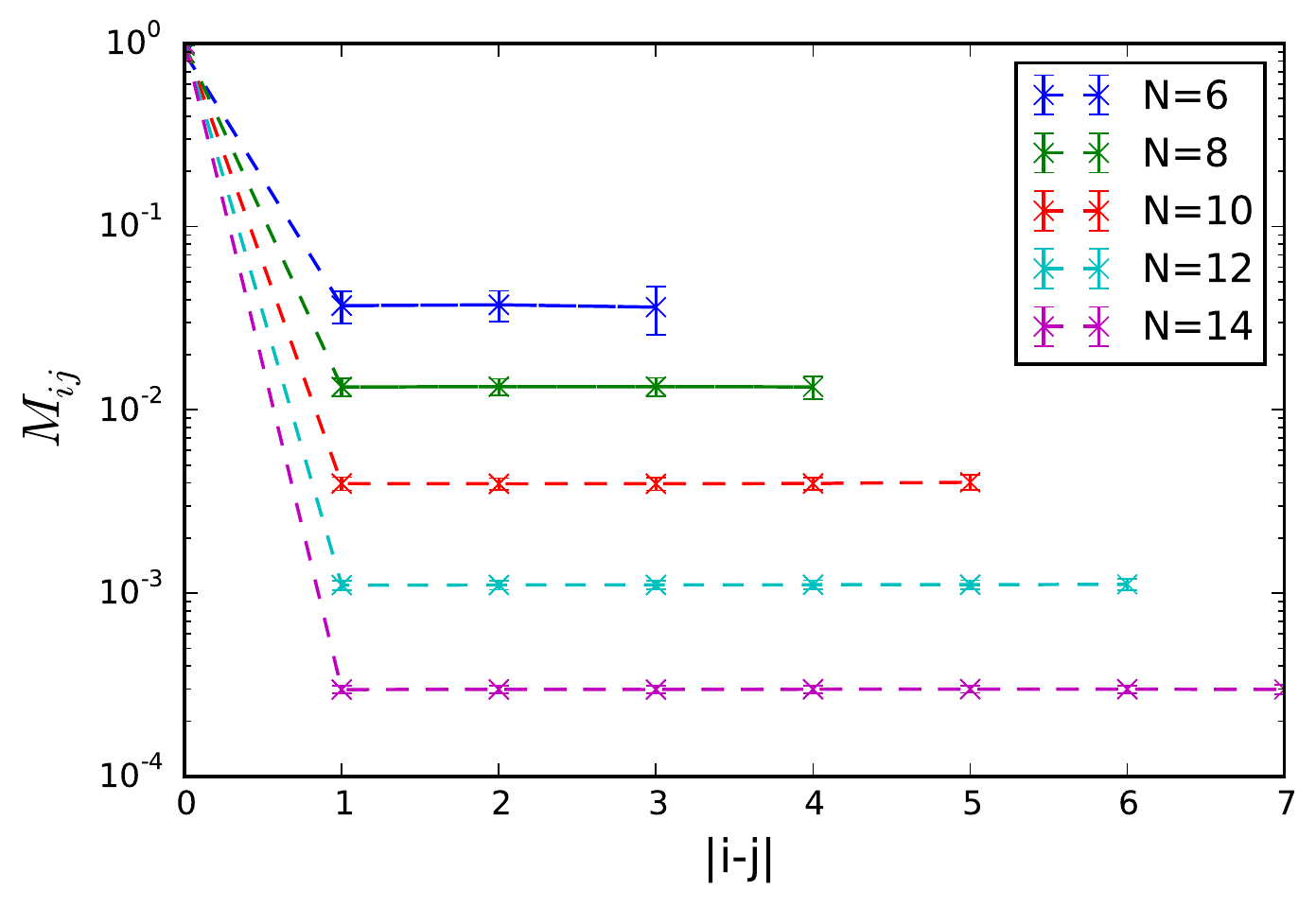}
\caption{The two point correlator for non-time averaged operators. This should be a delta function if $f=0$, and the corrections to that give the `background' plotted in the main text.}
\label{smallf_new}
\end{figure}

In Fig.~\ref{smallf_new} we show the `background operator' of Eq.~(\ref{background_eq}), but without the averaging over $j$. Such an operator would be a delta function in the limit $f\rightarrow 0$, we see instead that it is close to $1$ when $i=j$ and small and uniform otherwise. The uniformity at values $i\neq j$ justifies the averaging over $j$ which we performed in Eq.~(\ref{background_eq}).

\bibliography{lbits}

\end{document}